\documentclass[%
 amsmath,amssymb,
 aps,
 prb,
 twocolumn,
 superscriptaddress,
 showpacs,
 showkeys,
]{revtex4-1}

\usepackage{hyperref}
\usepackage{graphicx}
\usepackage{bm}
\usepackage{xcolor}
\usepackage{color}
\begin{document}

\title{Current-Induced Heat Transfer in Double-Layer Graphene} 

\author{Jiebin Peng}
\affiliation{Center for Phononics and Thermal Energy Science, School of Physical Science and Engineering, Tongji University, Shanghai 200092, China}
\author{Jian-Sheng Wang}
\affiliation{Department of Physics, National University of Singapore, Singapore 117551, Republic of Singapore}

\date{28 January 2019} 

\begin{abstract}
Using the fluctuational electrodynamics and nonequilibrium Green's function methods, we demonstrate the existence of a current-induced heat transfer in double-layer graphene even when the temperatures of the two sheets are the same. 
The heat flux is quadratically dependent on the current.
When temperatures are different, external voltage bias can reverse the direction of heat flow.
The drift effect can exist in both macroscopic and nanosized double-layer graphene and extend to any other 2D electron systems. These results pave the way for a different approach to the thermal management through radiation in 
nonequilibrium systems.
\end{abstract}
\maketitle 

\section{Introduction}
Understanding and controlling the heat flow is a significant endeavor both in nonequilibrium statistical physics and in 
practical applications. Managing radiative heat transfer (RHT) at small scales is essential for the development of a wide variety of technologies, including phononics \cite{bLi2012},  near-field thermophotovoltaics \cite{basu2009review} and thermal photonic analog of electronic devices \cite{PhysRevLett.112.044301}. 
In the last decades, near-field radiative heat transfer (NFRHT) \cite{song2015near}, where the separation distance is smaller than Wien's wavelength, has been proposed to enhance the RHT through surface-plasmon polariton \cite{joulain2005surface}, 
surface-phonon polariton \cite{RN14}, and so on. The NFRHT between different materials, such as semiconductor or bilayer graphene, can be electronically controlled by the photon chemical potential or gate voltage bias \cite{PhysRevB.91.134301,PhysRevB.85.155422,yu2017ultrafast,peng2015thermal}.  
Moreover, the novel electrically manipulated properties of 2D materials can give an additional knob to tune the RHT in nonequilibrium conditions. 

Here we explore the RHT between two graphene layers (double-layer graphene) across a separation gap by drifting one of the layers with a constant drift velocity or voltage bias, through modeling by the fluctuational electrodynamics (FE) and nonequilibrium Green's function (NEGF), respectively.  We demonstrate the existence of drift-induced RHT in double-layer graphene, with intensity depending quadratically on the drift velocities or voltage bias. The RHT produced by the temperature
imbalance can even be suppressed by drift-induced RHT, and the heat flux can be switched off by the voltage bias. We interpret that this drift effect is related to the negative Landau damping in graphene\cite{morgado2017negative}.   The physics is generic and it appears in between bulk graphene sheets or nanosized flecks. 

We consider a two-layer graphene system, where the bottom
and top layer are labeled by 1 and 2 respectively, separated by a vacuum gap of distance $d$, and emitting thermal radiation at temperatures $T_{1(2)}$. Figure 1 schematically illustrates the configuration under our investigation. There is an electric current induced by a static voltage applied across the bottom layer of graphene. Due to the high mobility of graphene, the drift velocity of electrons in graphene can be on the order of the Fermi velocity of graphene, i.e., $10^6\, \mathrm{m/s}$. 

\begin{figure}[htp] 
  \centering
  \includegraphics[totalheight=28mm]{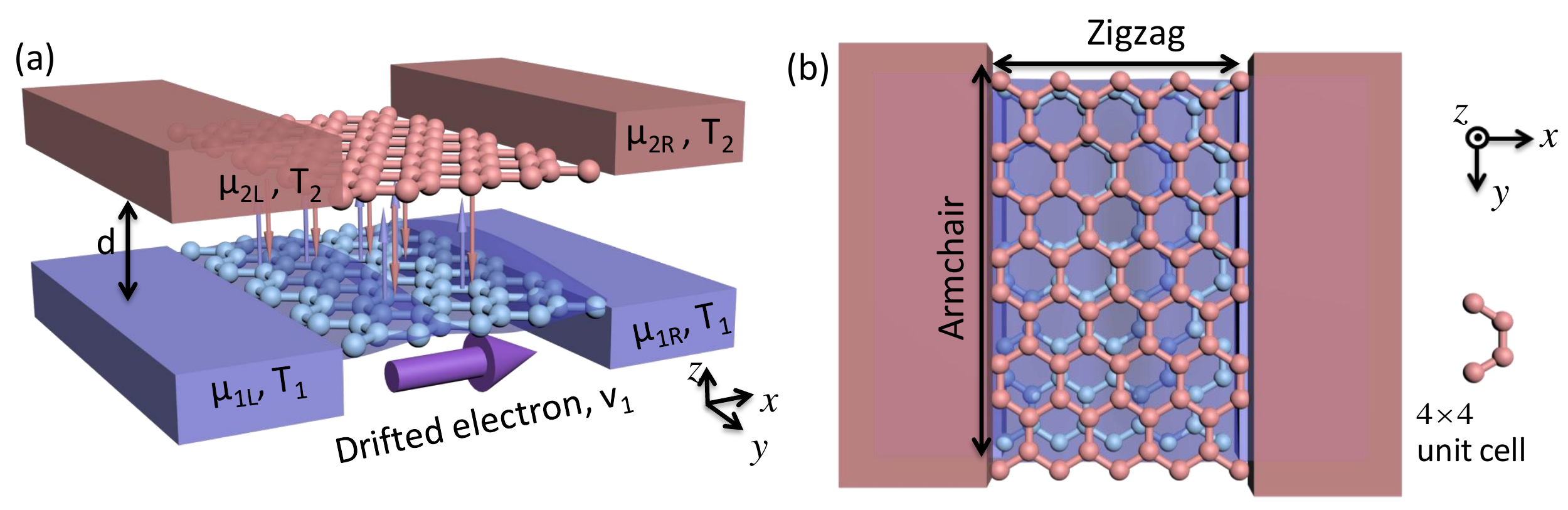}
  \caption{Scheme of a double-layer system under investigation: two graphene layers separated with a vacuum gap $d$.  The chemical potential difference ($\mu_{1R}-\mu_{1L}$) induces electron drift in the bottom graphene layer, (a) perspective view, (b) top view.}
 \label{figure1}
\end{figure} 

\section{Fluctuational electrodynamics description}

We generalize the usual formula for heat transfer, for which local equilibrium in each layer is assumed. 
Due to the current in the bottom layer, the system is intrinsically not in local thermal equilibrium.  This nonequilibrium 
situation is taken care by a simple hypothesis of an energy shift in the distribution functions.  When the bottom layer 1 is driven by a drift velocity $v_1$ in $x$ direction and there is no drift in the top layer, the photon distribution is Doppler shifted on the bottom layer.  Thus, the Bose function becomes:
\begin{align}
	N_1(\omega, k_x) = \frac{1}{e^{\frac{\hbar (\omega-k_xv_1)}{k_b T_1}}-1},
\end{align}
where $k_x$ is the wave number along $x$ direction.  $k_b$ is the Boltzmann constant.  Under the relaxation time approximation, the drift-induced distribution is similar to the distribution with chemical potential of photon in semiconductors.  
The heat transfer rate per unit area between the layers of graphene is then given under FE as \cite{rytov1959theory,RN7}: 
\begin{align} \label{RHT_general}
	H=\int_0^{\infty} \frac{d\omega}{2\pi}  \int \frac{d^2 {\bf k}_\perp}{(2\pi)^2} 
	\Theta_{12}(\omega, k_x) \mathrm{T}_{12} (\omega, {\bf k}_\perp),
\end{align}
where $\omega$ is the frequency of electromagnetic wave. ${\bf k}_\perp = (k_x, k_y)$ is the wavevector 
in the graphene plane. $\Theta_{12}(\omega,k_x) \equiv \hbar \omega\bigl[ N_1(\omega, k_x) - N_2(\omega) \bigr] $. 
$N_2(\omega)$ is the usual Bose function at temperature $T_2$. $\mathrm{T}_{12}(\omega,{\bf k}_\perp)$ is the transmission coefficient.
 
For an infinitely large suspended double-layer graphene with nanoscale separation, the $p$-polarized wave is the dominant channel for RHT \cite{PhysRevB.85.155422}. Based on FE, the transmission for evanescent $p$-polarized modes between 
a pair of two-dimensional materials in a parallel plate geometry can be written as
(in the non-retardation limit, the speed of light $c \to \infty$)
 \cite{polder1971theory,pendry1999radiative,RN7}:
\begin{align} \label{Transmisson_p}
	\mathrm{T}_{12} (\omega,{\bf k}_\perp)=\frac{4\,\mathrm{Im}(r_1)\, \mathrm{Im}(r_2)}{|1-r_1r_2e^{-2\gamma d}|^2} e^{-2\gamma d},
\end{align}
where $r_1$ and $r_2$ are the reflection coefficients at the bottom and top interface of the vacuum gap. 
$d$ is the distance of the vacuum gap. $\gamma = \sqrt{k_x^2+k_y^2} $.  
The drifted reflection coefficient is computed according to  
$r_1= v_{{\bf k}_\perp} \Pi(\omega, {\bf k}_\perp, v_1) / \bigl[1-  v_{{\bf k}_\perp} \Pi(\omega, {\bf k}_\perp, v_1)\bigr]$.
The bare Coulomb interaction in wavevector space in two dimensions is 
$v_{{\bf k}_\perp} = 1/(2\epsilon_0 \gamma)$. $r_2$ is computed at no drift ($v_1=0$). 
We calculate the drifted polarization function following the approximation of Svintsov et al.\cite{svintsov18}:
\begin{eqnarray}
\Pi_{{\bf k}_\perp}(\omega, v_1) &=& 
\frac{e^2 \mu(T_i)}{ (\pi \hbar v_F)^2} \int_0^{2\pi}\! d\theta\, 
\frac{1}{(1 - \beta \cos\theta)^2}\times \qquad \nonumber \\
&&  \frac{ k_x (\cos\theta - \beta) + k_y \sin\theta}{ (\hbar \omega + i \eta)/(\hbar v_F)
- k_x \cos\theta - k_y \sin\theta},\quad
\end{eqnarray}
where we define $\beta = v_1/v_F$, the Fermi velocity is $v_F = \frac{3}{2} a t/\hbar$ with carbon
bond length $a = 1.42\,$\AA\ and hopping parameter $t=2.8\,$eV,
and $\eta$ is a small electron damping parameter, which gives graphene a finite DC conductivity.
Finally, $\mu(T) = 2 k_b T \ln [ 2\cosh\frac{\mu}{2 k_b T} ]$. 
The long-wave approximation (${\bf k}_\perp$ small) is valid as the contribution of the transmission is concentrated around
$\gamma \sim O(1/d)$.   The appendices further discuss details of the approximations and calculations.


\begin{figure}[htp] 
  \centering
  \includegraphics[width=\columnwidth]{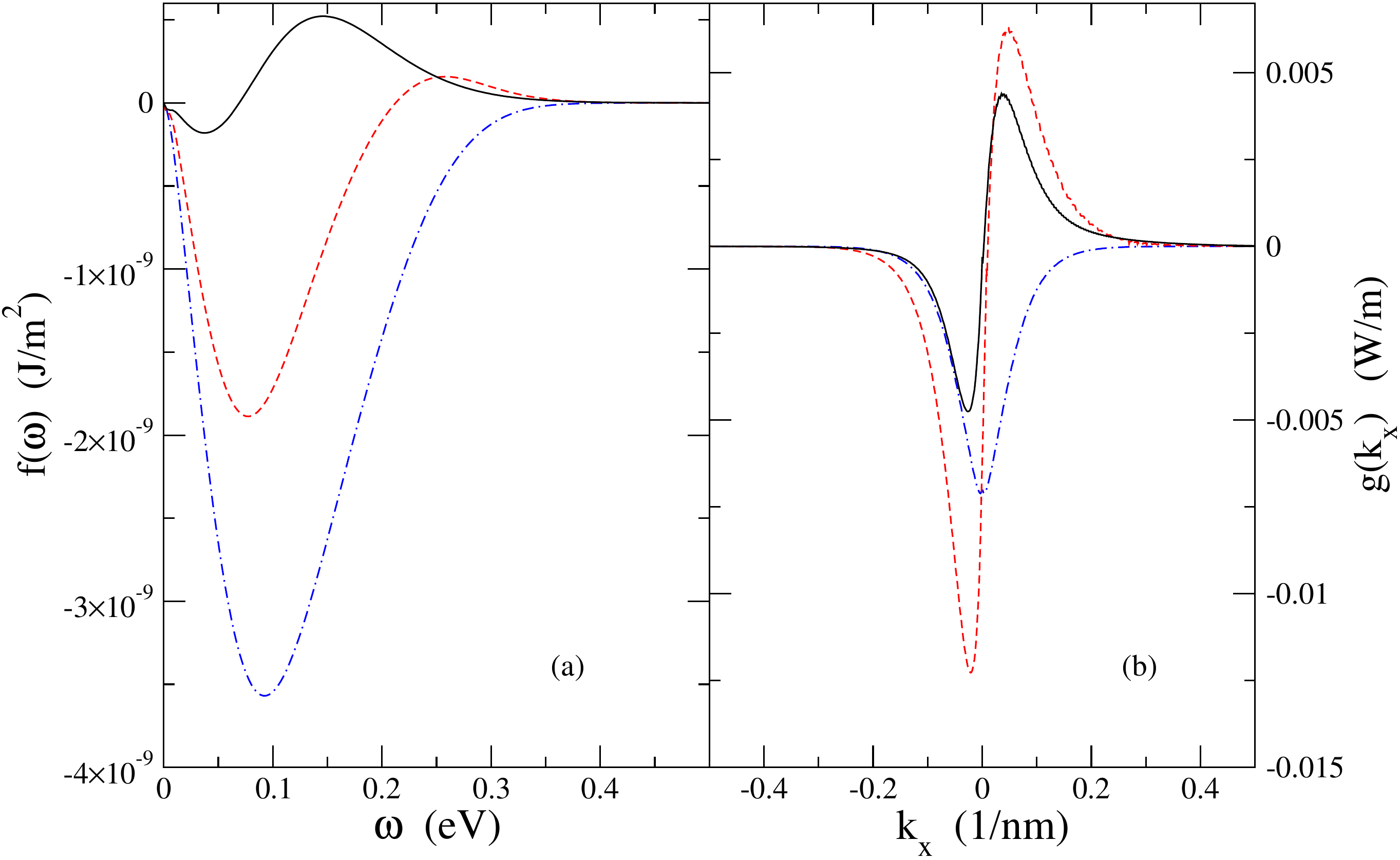}
  \caption{ (a) Integrated spectral transfer function $f(\omega)$ as a function of frequency and (b) 
  $g(k_x)$ as a function of wave vector, with different drift velocities: no drift (blue dash-dot line), total heat
  current density $H=-0.84\,$MW/m$^2$, $v_1=5.0 \times 10^5 \ \mathrm{m/s}$ (red dash line), $-0.30\,$MW/m$^2$, and $v_1=9.0 \times 10^5 \ \mathrm{m/s}$
  (black solid line), $+0.09\,$MW/m$^2$.  The temperatures are $T_1=300\,$K and $T_2=320\,$K. The chemical potential of graphene $\mu$ is set as 0.1\,eV. Gap distance $d$ is set as 10\,nm.  The damping parameter is $\eta = 9\,$meV.}
 \label{figure2}
\end{figure} 	

To understand the drift effects quantitatively, we define two integrated spectral transfer functions as follows:
\begin{align} \label{ISF}
	&f(\omega) =  \int \frac{d^2{\bf k}_\perp}{(2\pi)^3} \Theta_{12}(\omega,k_x)  \mathrm{T}_{12} (\omega,{\bf k}_\perp), \\
	&g(k_x)=\int_0^\infty \frac{d\omega}{2\pi} \int \frac{dk_y}{(2\pi)^2} \Theta_{12}(\omega,k_x)  \mathrm{T}_{12} (\omega, {\bf k}_\perp).
\end{align}
Figure~\ref{figure2}(a) shows the spectral transfer function $f(\omega)$ as a function of frequency with different drift velocities. In the undriven case (blue dash-dot), $f(\omega)$ is strictly negative and the heat current flows from top layer to bottom layer due to the temperature difference ($T_1 =300$\,K and $T_2=320$\,K). 
It is the usual RHT between two graphene layers and the RHT can be tuned by doping or top gating (varying $\mu$). However, when the electrons in the bottom layer are drifted with a velocity $v_1=5.0\times 10^5 \mathrm{m/s}$ 
(red dash line) in Fig.~\ref{figure2}(a), 
corresponding to an electric line current density $j_1 = (-e)n_1v_1 = 486$\,A/m, there is a small positive peak in $f(\omega)$. 
That means that part of the high-frequency modes can spontaneously emit external thermal radiation from bottom layer to top layer due to the drift velocity. Remarkably, with a higher drift velocity (black solid line), the height of the peak in the high-frequency region grows higher, and there is more heat transferred from bottom layer to the top layer.
Qualitatively, the heat flux generated by a temperature difference can be suppressed or even reversed by the high drift velocity.

The above mentioned drift-induced effects in suspended layers can be further understood in Fig.~\ref{figure2}(b): 
the distribution over the wavevector in the $\hat x$ (the driven) direction.  
In no drift case, $g(k_x)$ is negative and has the space inversion symmetry in $x$ direction.
However, when we drift the electrons in $x$ direction, the $k_x$ symmetry is broken and the drift induced mode appears. 
In the $k_x < \omega/{v_1}$ and $\omega>0$ region, it is negative and the system locates at the normal Landau damping region.
However, when $k_x > \omega/{v_1}$ and $\omega>0$, the drift induced modes carry positive value due to negative Landau damping. The total heat current is  from contributions of all those modes. In low frequency region ($\omega<0.1$\,eV), the heat flux is dominated by the normal Landau damping modes. 
In high frequency region ($\omega> 0.1$\,eV), the negative Landau damping modes can be the dominant modes for heat transfer. Due to the unsymmetric nature with an $x$-direction drift current, heat can be transferred from low temperature layer to high temperature layer through the negative Landau damping modes.  

\begin{figure}[htp] 
  \centering
  \includegraphics[width=0.85\columnwidth]{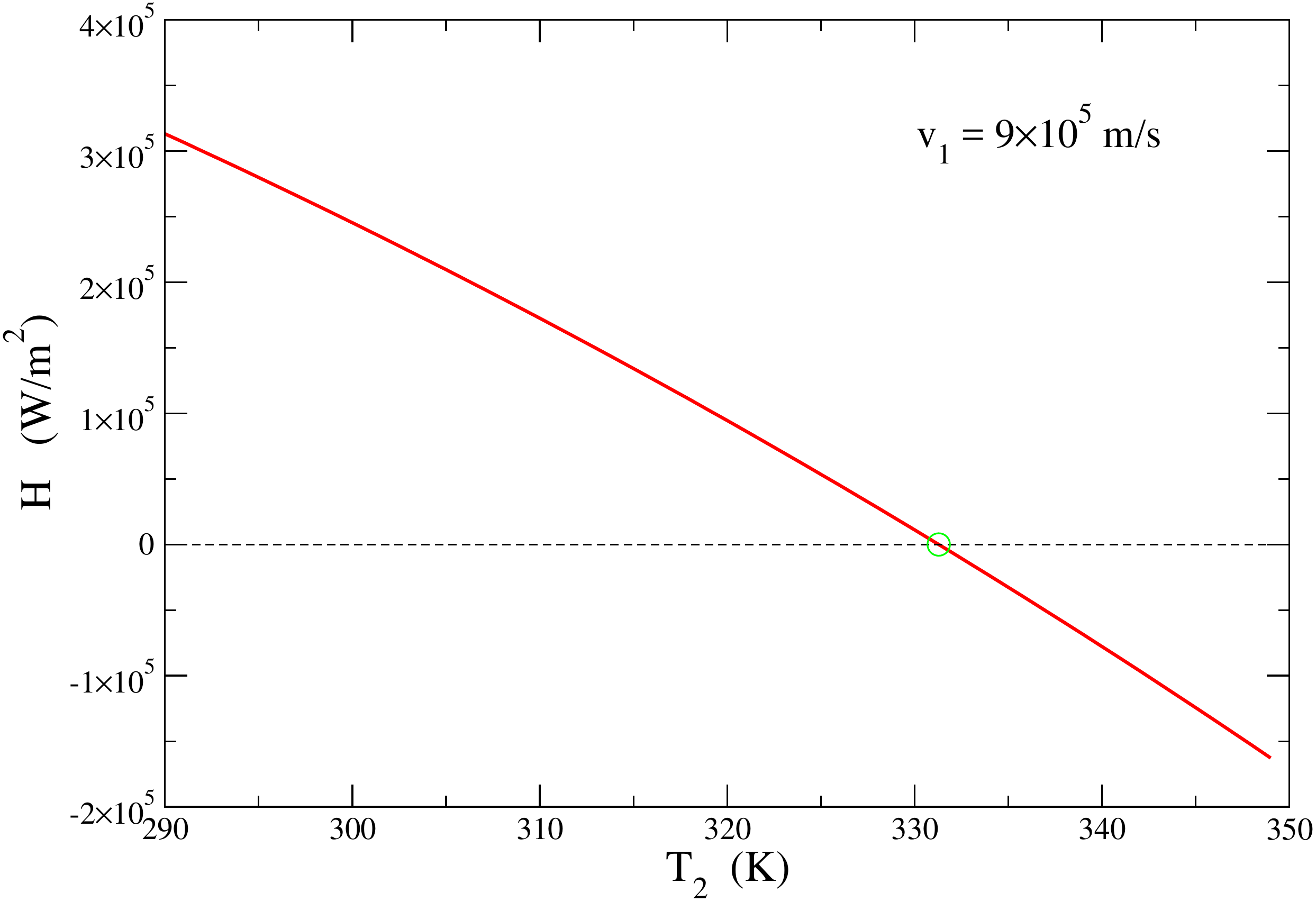} 
  \caption{Heat current density as a function of $T_2$ (temperature of the top layer) with drift velocity $v_1=9.0 \times 10^5 \ \mathrm{m/s}$ (red solid line).  Dotted line is the reference line for zero current density. The green circle indicates the point for ``off temperature''. The chemical potential of graphene $\mu$ is set as 0.1\,eV. Temperature of bottom layer of graphene $T_1$ is set as 300\,K. Vacuum gap distance $d$ is set as 10\,nm. }
 \label{figure3}
\end{figure} 

With the help of drift effects, we have demonstrated that the RHT in double-layer graphene can be tuned by the drift velocity, even shutting off or changing sign. To gain a detailed picture of the drift effects, we calculate the heat current density as a function of ${T_2}$ with fixed drift velocity $v_1=9.0\times 10^5$\,m/s. As seen in Fig.~\ref{figure3}, the heat current density is almost linearly decreasing as the temperature $T_2$ increases. When $T_2$ equals 331\,K, there is no heat current between the layers and the ``off temperature'' for RHT is reached. 


\section{Nonequilibrium Green's function calculation} 

The above calculation is based on FE for the infinite suspended double-layer graphene. To extend drift induced effect in the nanoscale system, we consider a small graphene nano-ribbon of
72 atoms in each layer (see Fig.~1), connected to two baths, in each layer.  
The chemical potential difference ($\mu_{1R}-\mu_{1L}$) between bath 1L and bath 1R produces the drift electrons in layer 1. In such a nanoscale system, the Coulomb interaction (virtual photon or scalar photon) will be the dominant mechanism for RHT. 

The calculation is based on a tight-binding model with a nearest neighbor hopping parameter $t=2.8$\,eV and Coulomb
interactions between the electrons \cite{zhang2018energy,wang2018coulomb,peng2017scalar}. 
The energy transfer out of layer 1 to layer 2 of a nanoscale double-layer graphene due to the Coulomb interaction 
can be calculated through the Meir-Wingreen formula  \cite{lu2016,lu-AIP-2015,MeirWingreen} under a lowest order expansion of
the Coulomb interaction (see Appendix~\ref{app-e} for a derivation):   
\begin{align}
H =- \frac{1}{A} \int_{0}^{+\infty} \frac{d\omega}{2\pi} \hbar \omega {\rm Tr} (D^> \Pi^<_1 - D^< \Pi^>_1 ). 
\end{align} 
Here $D^{>,<}$ is the greater/lesser Green's function for the scalar photon, which is calculated from the
Keldysh equation, 
$D^{<,>} = D^r \Pi^{<,>} D^a$, and retarded Green's function is obtained by
solving the Dyson equation, $D^r = v + v \Pi^r D^r$ ($v$ is the bare Coulomb potential with matrix element
$1/(4\pi\epsilon_0 r_{ij})$ between the tight-binding sites $i$ and $j$ of a distance $r_{ij}$).  
$\Pi^{<,>} = \Pi^{<,>}_1 + \Pi^{<,>}_2$ is block-diagonal and is obtained  with the random phase approximation (RPA).
$A$ is the area of the graphene.  Further computational details will be presented in  Appendix~\ref{app-d}.

Comparing with the calculation based on FE, the NEGF method provides a rigorous way to extend the drift effects into
nanoscale ballistic systems without any phenomenological assumptions. 
Due to the smallness of the sample, the transport is ballistic with the electric current at the bottom driven layer given
by $I_1 = (4 e/h) (\mu_{1R} - \mu_{1L})$, independent of the sample width or length.   We do not observe
Coulomb drag effect \cite{narozhny2016coulomb}, as the electric current in the top unbiased layer is very close to 0, while the thermal current going into it is quite large, see Fig.~4(b).  
From the view of energy transport, the energy can be transferred from an electrically driven layer to the closely spaced but electrically isolated layer: the drift electrons are dragged by the Coulomb interaction between two layers, and the energy can be transferred at the cost of the kinetic energy of drift electrons. 

\begin{figure}[htp] 
  \centering
  \includegraphics[width=\columnwidth]{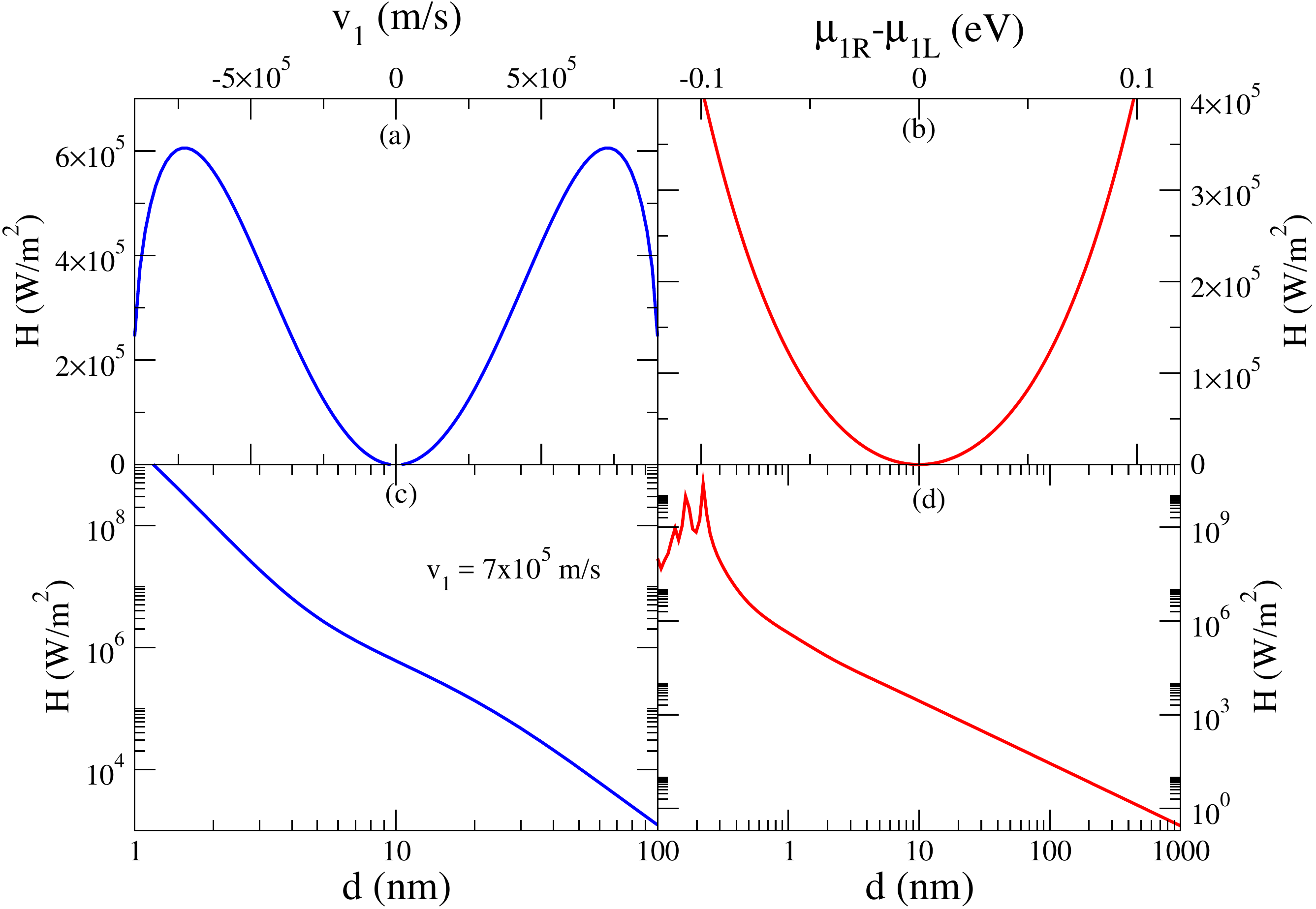}
  \caption{(a) FE calculation of drift-induced heat current density as a function of drift velocity. Vacuum gap distance $d$ is set as 10\,nm. The chemical potential of graphene $\mu$ is set as 0.1\,eV. (b) NEGF calculation of drift-induced heat current density as a function of chemical potential difference  $\mu_{1R}-\mu_{1L}$ symmetrically biased and $\mu_{2L}=\mu_{2R}=0.1\,\mathrm{eV}$. Vacuum gap distance $d$ is set as 1\,nm. (c) FE drift-induced heat current density as a function of gap distance with drift velocity $v_1=7.0 \times 10^5\,$m/s. (d) NEGF drift-induced heat current density as a function of gap distance with $\mu_{1R}=0.15$\,eV, $\mu_{1L}=0.05\,$eV and  $\mu_{2L}=\mu_{2R}=0.1\,$eV. There is no temperature difference between the double-layer graphene, $T_1=T_2=300\,$K.}
 \label{figure4}
\end{figure} 

Huge heat transfer appears even for $T_1 = T_2$, which is clearly due to strong Coulomb interactions 
at short distances. 
A comparison of FE and NEGF results is presented in Fig.~\ref{figure4}. Using FE method, Fig.~\ref{figure4}(a) indicates the drift-induced heat current density as a function of drift velocity with both layers at the same temperature. A parabolic dependence between drift-induced heat current and drift velocity is found numerically for small $v_1$.  For large
drift exceeding $|v_1| > 7.0\times 10^5\,$m/s, we see non-monotonic behavior. 
From the NEGF point of view, the chemical potential difference between left and right bath can produce an 
electric current and it is similar to the drift electron one in FE calculation. 
In Fig.~\ref{figure4}(b), we found that the heat current density also quadratically  depends on the chemical potential difference $\mu_{1R}-\mu_{1L}$.  
In our symmetric setup, we do not expect that the heat transfer will be different if we reverse the direction of
the drift velocity or voltage bias, so it must depend on them quadratically. 
It is tempting to assume that the heat transfer is given by Joule heating caused by Coulomb drag.   Unfortunately,
this is not quite true  (see Appendix~\ref{app-c}).

Further numerical evidence is shown in Fig.~\ref{figure4}(c) and (d): the distance dependence of drift-induced heat current.  Fig.~\ref{figure4}(c) depicts the drift-induced heat current density as a function of distance under FE calculation with fixed drift velocity.
We observe that the drift induced heat current decays initially as $d^{-4}$ when the gap distance is smaller than 10\,nm 
and decreases as $d^{-2}$ when $d>10$\,nm. 
A similar distance dependence is also observed in Fig.~\ref{figure4}(d) for the nanosize sample with a 
larger range of $d^{-2}$ behavior.

In summary, we proposed methods to describe heat transport without the assumption of local equilibrium.   With FE
method, the Bose function and Fermi function need to be shifted due to the drift velocity.  For NEGF, we give
a Meir-Wingreen formula with the input from a RPA calculation for $\Pi^r$ where the electron Green's function is  current-carrying and not in thermal equilibrium.  We have demonstrated a drift induced radiative heat transfer in double-layer graphene. 
Such effects can be extended to any other 2D electron systems with large drift velocity or current. 
The drift induced heat
current can even shut off the heat current produced by a temperature difference. It enables possibilities to exploit 
RHT through electronic control.  Further, the proposed drift effects can exist in both large and microscopical systems. 
With wide-band tunability and nano-scale characteristic dimensions, the proposed drift effects appear very charming 
for the application in radiative thermal management.  

\section*{Acknowledgements}
We thank Han Hoe Yap for comments, and Jia-Huei Jiang for the codes.
This work is supported by FRC grant R-144-000-402-114 and MOE grant MOE2018-T2-1-096.


\appendix

\section{Doppler shift}
In the appendices, we clarify some points made in the main texts and give some further 
details.

We first note that the function $\Pi^r_{\bf k}(\omega)$ is a description of the bosonic charge density plasmon.  
In particular, we consider a plasmon planewave $e^{i( {\bf k} \cdot {\bf r}-\omega t)}$ with frequency $\omega$ and 
wavevector ${\bf k}$.  Let ${\bf v}_1$ describe the drift velocity of the electrons.    If the wave ${\bf k}$
of the plasmon and the electron drift velocity ${\bf v}_1$ are in the same direction, electrons will
``see'' less vibrations, thus $\omega$ decreases.   So the Doppler shift is,
$\omega \rightarrow \omega + \delta \omega =  \omega - {\bf k} \cdot {\bf v}_1 = \omega - k_x v_1$.

\section{Doppler shift or not for scalar photon self-energy}

The scalar photon self energies or polarization functions under the random phase approximation are, 
in time domain and real space,
\begin{eqnarray}
\Pi_{jk}^>(t) &=& (-i\hbar)e^2 G_{jk}^>(t) G_{kj}^<(-t),\\
\Pi_{jk}^<(t) &=& (-i\hbar)e^2 G_{jk}^<(t) G_{kj}^>(-t),\\
\Pi_{jk}^r(t) &=& \theta(t) \bigl( \Pi_{jk}^>(t) - \Pi_{jk}^<(t)  \bigr).
\end{eqnarray}
We obtain the frequency and wavevector domain quantities if we Fourier transform the time and space.   For simplicity
of notation, we consider spinless electrons with a single band, $\epsilon_{\bf k}$.  Since the system is not
in thermal equilibrium, we cannot evoke the usual fluctuation-dissipation theorem, but something very close to it.  
We use the Kadanoff-Baym ansatz \cite{kadanoffbaym}, i.e.,
\begin{equation}
G^< = -f (G^r - G^a),\quad  G^> = (1 -f) (G^r - G^a),
\end{equation}
where $f$ is given by the solution of Boltzmann equation.  The spectrum function, $A = i (G^r - G^a)$, is assumed to 
be evaluated in thermal equilibrium.   In wavevector and angular frequency domain, we can write \cite{mahan00,duppen16}
\begin{equation}
\label{eqpir}
\Pi^r_{\bf k}(\omega) = - e^2 \int \frac{d^2\bf p}{(2\pi)^2} \frac{ f_{\bf p} - f_{{\bf p} - {\bf k}}}{\hbar \omega + i \eta - 
\epsilon_{\bf p} + \epsilon_{{\bf p} - {\bf k}}}.
\end{equation}
Here the integration is over the first Brillouin zone, and $\eta$ is a small damping parameter inversely proportional to the relaxation time. 

As the relaxation mechanism for the electrons can be very complicated due to different
scattering possibilities -- impurity scatterings, electron-electron and electron-phonon scatterings -- we do not
attempt to solve the Boltzmann equation, and just use a single-mode relaxation time approximation \cite{ziman60}.  In such a
framework, we can write
\begin{equation}
f = f^0 - \frac{df^0}{d \epsilon} \Phi \approx f^0(\epsilon - \Phi),
\end{equation}
here $\Phi \equiv \Phi_{\bf k}$ is mode ${\bf k}$ dependent, and $f^0 = 1/\bigl[\exp((\epsilon - \mu)/(k_bT)) + 1\bigr]$
is the equilibrium Fermi distribution at temperature $T$ and chemical potential $\mu$.

The effect of the current drift is to introduce anisotropy to the problem, thus we expect $\Phi$ should be angle
and magnitude dependent, $\Phi_{\bf k} = \Phi(\theta, k)$, here $\theta$ is the angle between ${\bf k}$ and
the drift velocity ${\bf v}_1$ and $k = | {\bf k}|$ is the magnitude of the wavevector.  We make a Legendre
polynomial expansion of the angular dependence and keep only the lowest non-trivial term, i.e., we write,
$\Phi \propto \cos(\theta)$.  
It is convenient to assume
\begin{equation}
\label{eqphik}
\Phi_{\bf k} =  \hbar k_x v_1 =  \hbar v_1 k \cos\theta. 
\end{equation}
If $\Phi$ does take this linear dependence on $k_x$, the nonequilibrium distribution
might be transformed back to the equilibrium one of $f^0$ by a change of reference frame.
That is, 
\begin{equation}
\label{noneq-eq}
\Pi^{>,<,r}_{{\bf k}, \rm noneq}(\omega) \approx \Pi^{>,<,r}_{{\bf k}, \rm eq}(\omega - k_x v_1).
\end{equation}
In steady state, the fluctuation-dissipation like relation for photon self-energy can be obtained from Eq.~(\ref{noneq-eq}) even if there is drifted electron current:
\begin{align}
	&\Pi^{<}_{{\bf k}, \rm noneq}(\omega) = 2 i N_{{\bf k}, \rm noneq}(\omega) \mathrm{Im} \Pi^{r}_{{\bf k}, \rm noneq}(\omega),\\
	&N_{{\bf k}, \rm noneq}(\omega)\approx N(\omega-k_x v_1),
\end{align}
where $N(\omega) = 1/(e^{\hbar \omega/(k_bT)}-1)$ is the Bose distribution.

The reflection coefficient needed for the heat transfer calculation is then obtained from the relation
$r = v \Pi^r/( 1 - v \Pi^r)$ with bare Coulomb potential in two dimensions $v = 1/(2 \epsilon_0 k)$.
The optical conductivity is related to the retarded self energy by
$\sigma = i \frac{\omega}{k^2} \Pi^r_{\bf k}(\omega)$.


\begin{figure}[htp] 
  \centering
  \includegraphics[width=\columnwidth]{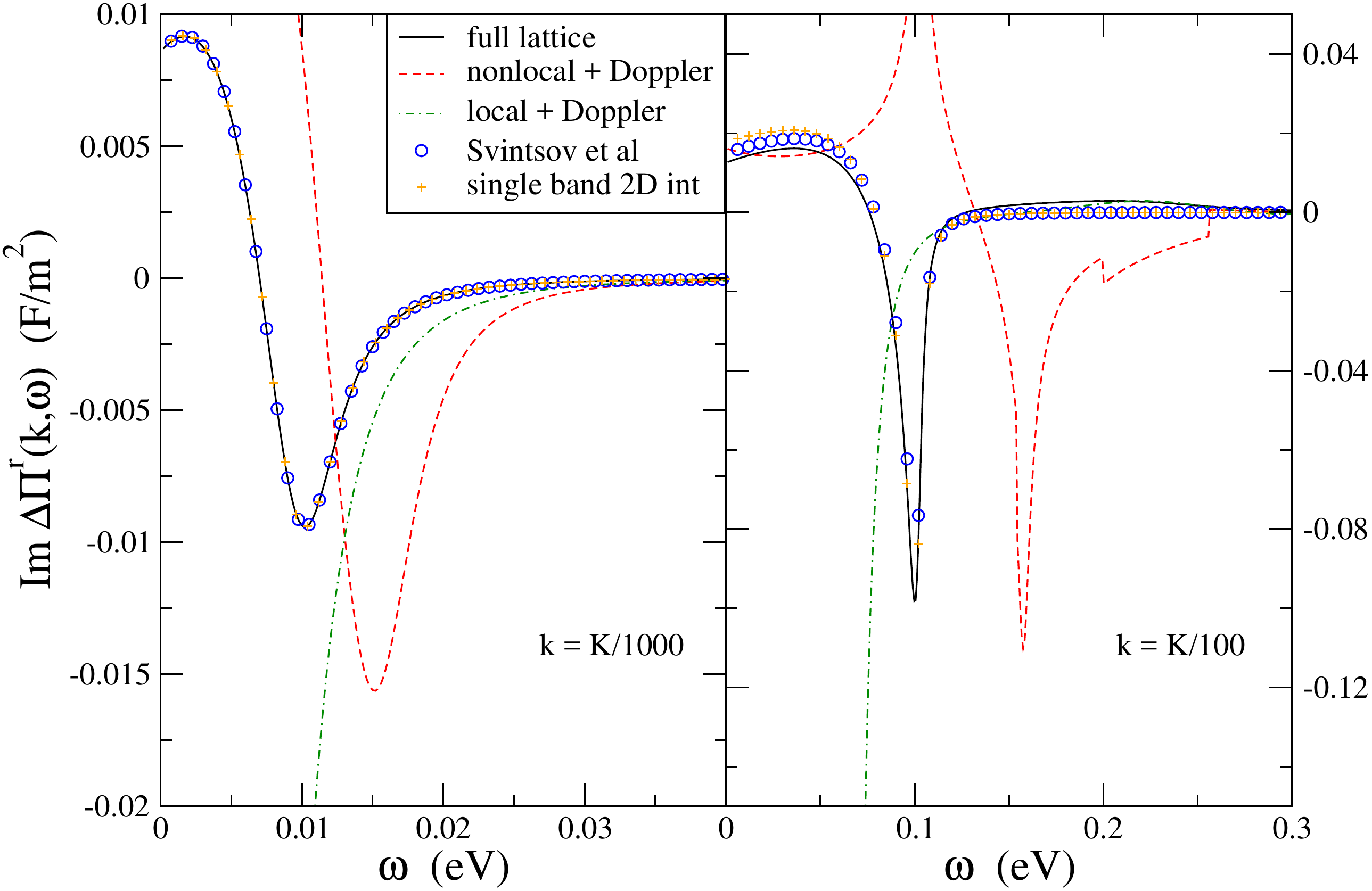}
  \caption{Imaginary part of $\Pi^r({\bf k}, \omega, v_1) - \Pi^r({\bf k}, \omega, 0)$, with 
  ${\bf k} = (k_x,0)$, $v_1 = 5 \times 10^5$\,m/s, at chemical potential $\mu = 0.1\,$eV.  Left graph has
  $k_x = K/1000$ and right 10 time larger, $K = 4\pi/(3 \sqrt{3} a)$ is the magnitude of the K-point vector from
  the $\Gamma$ point, $a=1.42\,$\AA\ is carbon bond length. Temperature is set at 100\,K.   
  A damping parameter in $\hbar \omega \to \hbar \omega + i\eta$ is set to $\eta  = 3.3\,$meV.
  The legends apply to both graphs,  full black solid line:
  tight-binding lattice model with nearest neighbor hopping parameter $t=2.8\,$eV, 
  red dash line: Doppler shifted Wunsch et al.\ expression, green dot-dash line: Doppler shifted Falkovsky's expression,
  blue circles: Svintsov et al expression, orange pluses: numerical integration of Eq.~(\ref{eqpir}).} 
  \label{figurePi}
\end{figure} 

For a quadratic dispersion relation, $\epsilon_{\bf k} = \hbar^2 k^2/(2m)$, the claim, Eq.~(\ref{noneq-eq}), is easily verified using the explicit expression for $\Pi^r$, Eq.~(\ref{eqpir}), by a change of integration variable with a constant shift,
${\bf p} \to {\bf p} + m{\bf v}_1/\hbar$.  
For graphene with Dirac cone, $\epsilon_{\bf k} = v_F\, \hbar k$, this is no longer true.  
A variable transform cannot eliminate both $\Phi_{\bf p}$ and $\Phi_{{\bf p}-{\bf q}}$ simultaneously.
As a result, Doppler shift of the equilibrium result and nonequilibrium distribution in the Fermi function
becomes inequivalent\cite{svintsov18}. The energy shift
needs to be momentum dependent.
The linear dependence on $k_x$ times a constant requires a specific assumption on the energy dependence
of the relaxation times.  According to the usual relaxation time approximation, $\Phi$ is given by (near
$K$ or $K'$ points)
\begin{equation}
\Phi = - e E \tau_{k} \frac{\partial \epsilon_{\bf k}}{\partial ( \hbar k_x )},
\end{equation}
where electron carries charge $-e$, $E$ is the applied electric field in $x$ direction, 
$\tau_k$ is the relaxation time assumed to only depend
on the magnitude of ${\bf k}$, and the last factor is the group velocity.   For a metal with the usual quadratic
dispersion, Eq.~(\ref{eqphik}) is true if we use constant relaxation time, which turns out to be very good
approximation for metal.   For graphene, the group velocity in $x$-direction is $v_F \cos(\theta)$ with
the Fermi velocity a constant, thus we
must demand a relaxation time proportional to $k$, which turns out in agreement with experiments
\cite{antonio-RMP2009}.

For a full lattice model numerical calculation such that $\Phi_{\bf k}$ respects the lattice symmetry, we take the drift term to be
$\Phi_{\bf k} = v_1 {\rm Re} \left[ z^{*} \frac{\partial z}{\partial k_x} \right]/ (\hbar v_F^2)$, where
$v_F = 3 a t/(2\hbar)$ is the graphene Fermi velocity, and $z = 
-t ( e^{-i k_x a} + e^{i (k_x a/2 + k_y a \sqrt{3}/2)} + e^{i (k_x a/2 - k_y a \sqrt{3}/2)} )$.
 In Fig.~\ref{figurePi}, we compare Doppler shifted
Wunsch et al.\ expression \cite{guinea06} (or that of Hwang and Sarma \cite{hwang-sarma07}) which is a nonlocal result 
at zero temperature or that of Falkovsky's small
$q$ (local) result\cite{falkovsky08} at $T=100\,$K.  As we can see they differ a lot and do not agree with the correct way of
evaluating the drifted polarization function.  On this scale of vertical axis, they also tend to
diverge to plus or minus infinity.   However, Svintsov et al.\ expression \cite{svintsov18} (with a correction 
of a sign error in the denominator), 
\begin{eqnarray}
\Pi^r_{(k_x, k_y \!=\! 0)}(\omega, v_1) &=& 
-\frac{2}{\pi} \frac{e^2 \mu}{(\hbar v_F)^2} \frac{1}{(1-s\beta)^2}\Bigg(\qquad\qquad\qquad \\
 &&\qquad \sqrt{1-\beta^2}  - \frac{s-\beta}{\sqrt{s^2-1}}\Bigg), \\
\beta &=& {\rm sgn}(k_x) \frac{v_1}{v_F}, \quad s = \frac{\hbar \omega + i\eta}{\hbar |k_x| v_F},
\end{eqnarray}
agrees very well with a full lattice model calculation \cite{jia-huei17}.   When the chemical potential is much larger than
$k_b T$ and at $\omega > v_F k \to 0$, we have an excellent approximation for the equilibrium polarization
\begin{equation}
\Pi^r_{\bf k}(\omega) \approx \frac{e^2 \mu}{\pi} \frac{k^2}{(\hbar \omega + i \eta)^2}.
\end{equation}
Doppler shift of this expression  in the small $v_1$ limit gives $\Delta \Pi^r = \Pi^r(v_1) - \Pi^r(0) \approx
2 D$, while Svintsov et al.\ expression in this limit is $-D$, where $D= (e^2 \mu/\pi)\hbar v_1 [k/(\hbar \omega + i \eta)]^3$.   Both the sign and magnitude are different, contrary to the claim in ref.~\onlinecite{morgado-reply18}.

\section{Possible connection to Coulomb drag\label{app-c}}
Since we expect that the drift velocity $v_1$ is a small quantity, we can make linear, or rather, quadratic response
calculations.   Taylor expanding in variable $\delta \omega = - k_x v_1$ of the distribution $N_1$ of Eq.~(1) in the
main text to second order and the transmission function to first order, 
and then substituting into Eq.~(2), and setting the temperatures $T_1 = T_2 = T$, we obtain
\begin{eqnarray}
H &=&  \int_{-\infty}^{+\infty}\!\! \frac{d\omega}{4\pi} \hbar \omega 
\int \frac{d^2 {\bf k}}{(2\pi)^2}  T_{12}({\bf k}, \omega)\Big|_{v_1=0} \frac{1}{2} \frac{\partial^2 N}{\partial \omega^2} 
\bigl( \delta  \omega \bigr)^2 \nonumber \\
&+& \int_{-\infty}^{+\infty}\!\! \frac{d\omega}{4\pi} \hbar \omega 
\int \frac{d^2 {\bf k}}{(2\pi)^2}  \frac{\partial T_{12}({\bf k}, \omega)}{\partial v_1}\Big|_{v_1 =0}   \frac{\partial N}{\partial \omega} 
\bigl( v_1 \delta  \omega \bigr) \nonumber.
\end{eqnarray}
The first order term proportional to $\delta \omega$ is 0 because it is an odd function in $k_x$. 
We will write this long expression as $H = a\, v_1^2 = (a_1 +a_2) v_1^2$ where $a_1$ is from the $(\delta \omega)^2$ 
term and $a_2$ from the cross term.  The coefficient $a_2$ is complicated, hence we focus only on $a_1$.
Since the integral over $k_x^2$
and $k_y^2$ factor is the same, we symmetrize the formula about $x$ and $y$ and divide by two. With further
simplification of the second derivation of the Bose function, we obtain
\begin{equation}
a_1 = \int_{-\infty}^{+\infty} \frac{d\omega}{4\pi} \int \frac{d^2 {\bf k}}{(2\pi)^2}
 \frac{ T_{12}({\bf k}, \omega)\, k^2 (\beta \hbar)^2 \hbar \omega }{ 16 \sinh^2(\frac{\beta \hbar \omega}{2}) 
\tanh(\frac{\beta \hbar \omega}{2}) }.
\end{equation}
Here $T_{12}$ is defined in the main text by Eq.~(\ref{Transmisson_p}), evaluated at $v_1=0$, 
$\beta = 1/(k_b T)$.   Finally, we make one approximation, that is to assume
$\tanh(x) \approx x$ valid if frequency is small in comparison with temperature.  Then, we find
\begin{equation}
a_1 \approx e^2 n_1^2 \rho_D,
\end{equation}
if we compare our formula with that of Jauho and Smith\cite{jauho-smith1993} (Eq.~(5) and (27)) and that of 
Flensberg and Hu\cite{flensberg-hu1995} (Eq.~(2) and (20)).  Here $n_1$ is the carrier 
surface density and $\rho_D$ is the Coulomb drag coefficient.     Because of the existence of the second
term, $a_2$, we don't have a simple interpretation of Joule heating due to Coulomb drag.  Numerically,
for the parameters used for Fig.~4(a) in the main texts, we find $a_1 = 5.5 \times 10^{-6}\,$Ws$^2$/m$^4$,
while $a_2 = - 4.3 \times 10^{-6}\,$Ws$^2$/m$^4$.  There is a cancellation effect, given a smaller overall
$a$. 


\section{Numerical calculation of the NEGF conjunction system\label{app-e}}
For the 4-terminal junction double-layer system discussed in the main texts, wave-vector is not a good quantum number
and we cannot study finite size transport in ${\bf k}$ space.   As a result, we do calculation in real space with the electron 
Green's functions $G_{jk}^{>,<,r}$ where $j$ and $k$ runs over the sites of top and bottom layers of graphene.
The retarded Green's function is calculated in energy space with $G^r(E) = ( E I - H_C - \Sigma^r)^{-1}$ where
$\Sigma^r$ is the sum of total self energies due to the four leads.  Actually, both $H_C$ and $\Sigma^r$  are
block diagonal with respect to the layer index since there is no direct electronic coupling.  As a result, $G^{>,<,r}$ is also
block diagonal.  The lead self energies are calculated by standard iterative algorithms of surface Green's functions.

The polarization functions $\Pi^{>,<,r}$ are calculated according to the time-domain formulas and then Fourier
transformed to frequency domain.   This appears very fast computationally, but it brings about numerical instability
for large systems.  $4 \times 4$ system with about 200 atoms is probably the largest system we can obtain
reliable result from.    We have used 23000 fast Fourier transform points with a spacing $\hbar \Delta \omega =
22$\,meV.
The nonequilibrium information is incorporated through the lead temperatures and chemical potentials by 
the Keldysh equation, $G^< = G^r \Sigma^< G^a$, at the very beginning when calculating the polarization function
$\Pi^{>,<}$, and not through somewhat ad hoc procedure such as Doppler shifting the Bose function. 
$D^r$ is obtained with the Dyson equation, $D^r = v + v \Pi^r D^r$, and $D^{>,<}$ is obtained by the
corresponding Keldysh equation.   Since we are in real space, we have used periodic boundary conditions
in the $y$ direction perpendicular to the transport direction, and have set the diagonal $v_{ii} = 0$. 
Finally, the heat current is calculated according to the formula given in the main texts, Eq.~(7).  The electric current
can also be calculated, using the lowest order expansion formula for the interacting Green's function given
in the next section, with the Meir-Wingreen formula for electric current.

\section{Proof of the scalar photon Meir-Wingreen formula, Eq.~(7)\label{app-d}}

We give a derivation of the Meir-Wingreen formula for scalar photon
in terms of the Meir-Wingreen formula for the electrons under the lowest order of expansion 
approximation \cite{paulsson2005,lu2016}.   
We consider a two-layer setup with four leads,
layer 1 with left and right lead, and layer 2 left and right lead.  The energy current out the layer $\alpha$
is \cite{MeirWingreen,lu-AIP-2015}
\begin{equation}
\label{MWeq1}
I_\alpha = \int_{-\infty}^{+\infty} \frac{dE}{2\pi \hbar} E\, {\rm Tr} \bigl( G^> \Sigma_\alpha^< 
- G^< \Sigma_\alpha^> \bigr).
\end{equation}
Here $G^>$ and $G^<$ are the full interacting greater and lesser Green's functions of the electrons and 
$\Sigma_{\alpha}^{>,<} = \Sigma_{\alpha,L}^{>,<} + \Sigma_{\alpha,R}^{>,<}$ are the total
lead self energies.  They are functions of energy $E$.
This formula is exact provided that the electron Green's function is obtained exactly.  However, such a goal 
for the Coulomb system is not attainable.   Thus, we use the lowest order expansion approximation in
terms of the Coulomb interaction.  Such approximation preserves energy conservation exactly.  Since
the two layers are not coupled directly, the Green's functions for the electrons and self energies 
are block diagonal, and the Meir-Wingreen formula needs only the block $\alpha$.  
We focus on layer 1, and Green's function $G_{1}^>$, can be expressed by the Keldysh equation as
\begin{equation}
G_1^> = \bigl[ G^r( \Sigma_1^> +\Sigma_2^> + \Sigma_n^> ) G^a \bigr]_{11}.
\end{equation}
Here $\Sigma_{1,2}^>$ are the lead self energies, and $\Sigma_n^>$ is the Fock term of Coulomb
interaction, all of them block diagonal.
For the 11-subblock, $\Sigma_2$ is 0.  Putting this result into the Meir-Wingreen formula, noting
that ${\rm Tr}( G_1^> \Sigma_1^< - G_1^< \Sigma_1^>)=0$ as a consequence of charge and energy
conservation when Coulomb interaction is turned off, we obtain
\begin{equation}
I_1 = \int_{-\infty}^{+\infty} \frac{dE}{2\pi \hbar} E\, {\rm Tr} \bigl[ G^r_1 \Sigma_n^> G_1^a 
\Sigma_1^< - ({\rm swap\ } {}^> \leftrightarrow {}^< ) \big]. 
\end{equation}
From now on we will drop the subscript 1 for notational simplicity. 

A key approximation we use is the lowest order expansion,
\begin{equation}
\label{LOEeq}
G^> \approx G_0^> + G_0^r \Sigma_n^r G_0^> + G_0^r \Sigma_n^> G_0^a + G_0^> \Sigma_n^a G_0^a.
\end{equation}
We obtain such terms if we expand the contour ordered Dyson equation, $G = G_0 + G_0 \Sigma_n G
\approx G_0 + G_0 \Sigma_n G_0 + \cdots$, and then take the greater component using the 
Langreth rule \cite{haug-jauho-book}.  We also drop the subscript 0 from now on.

\begin{figure}[htp] 
  \centering
  \includegraphics[totalheight=55mm]{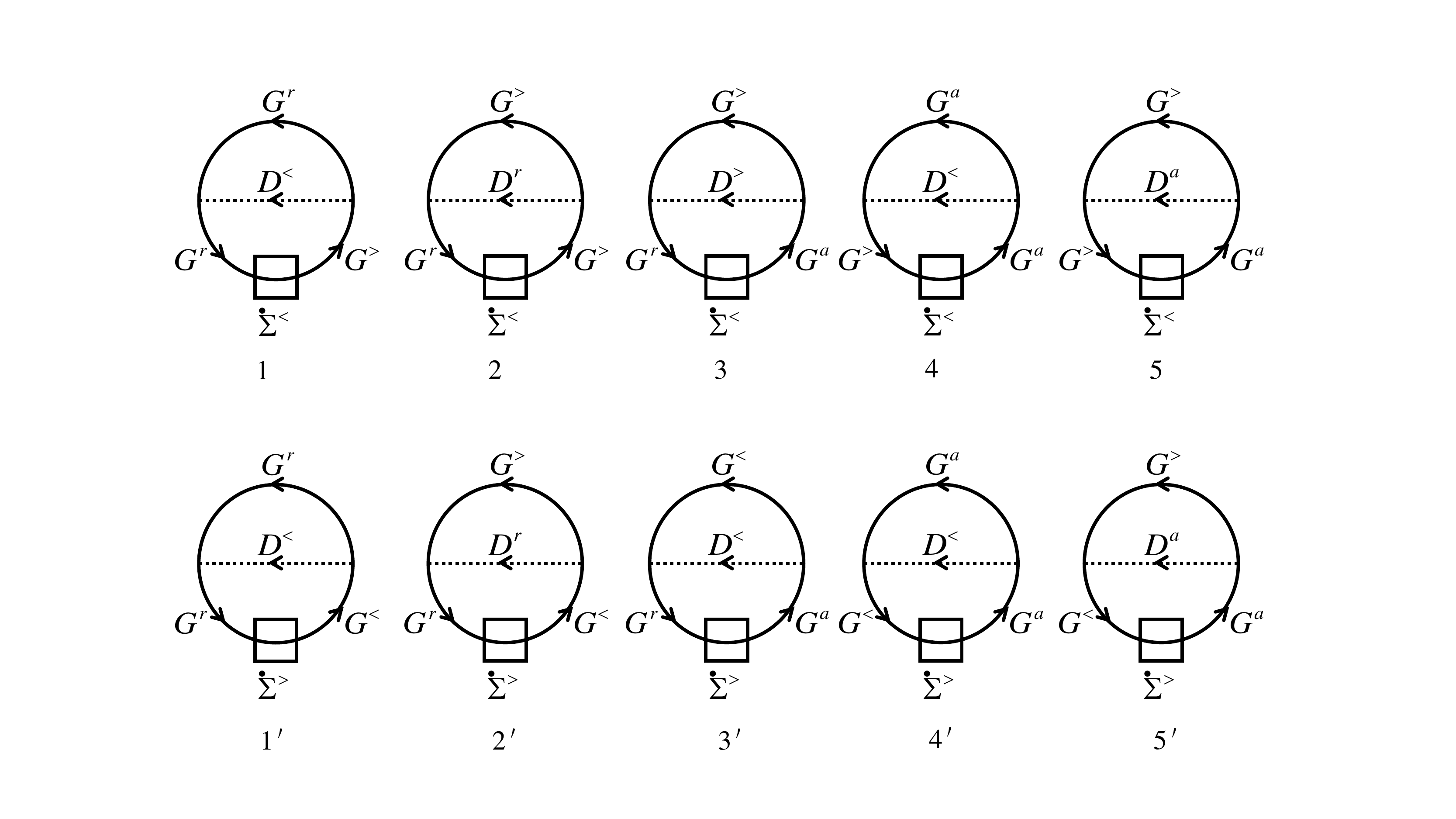}
  \caption{Diagrams for heat current in lowest order expansion.}
 \label{feynman-Diagrams}
\end{figure} 

It is useful for symmetry reasons we express the current by vacuum diagrams in time domain.  
We use the inverse Fourier transform to change the integral in energy to time, and also the Fock diagram result, $\Sigma_n^>(t,t') = i \hbar \sum_{l,l'}
M^l G^>(t,t') M^{l'} D_{l,l'}^>(t,t')$.  Similar expressions are given for retarded and advanced
self energies as $\Sigma_n^r \propto D^< G^r + D^r G^>$, and $\Sigma_n^a \propto 
D^< G^a + D^a G^>$.
Here for generality, we assume the interaction bare vertex
takes the form $\sum_{l} c^+ M^l c \phi_l$, where $c$ is a column vector of electron annihilation operators 
and $c^\dagger$ is row vector of hermitian conjugate, and $\phi_l$ is scalar field at site $l$, and $M^l$ is a hermitian matrix.   A matrix multiplication,
$M G M$, is implied in the electron space index.   By plugging in Eq.~(\ref{LOEeq}) into (\ref{MWeq1}), 
the expansion leads to 10 terms, represented
by the 10 diagrams in Figure \ref{feynman-Diagrams}.  We will label these diagrams as 1 to $5$, and $1'$ to $5'$ as shown. The diagrammatic rule follows the usual convention with all the (real) times
as dummy integration variables and space indices summed.   The current is obtained by
$(i\hbar)^2/T$ times the value of the diagram.  Since all the times are integration variables on equal footing, 
the integral actually diverges, the $1/T$ factor cancels the last integral interpreted as $\int_{-T/2}^{T/2} dt \cdots$.   As 
an example, the graph 3 represents the contribution to current as
\begin{eqnarray}
3) &=& \frac{(i\hbar)^2}{T} \int dt dt' dt_1 dt_2 \sum_{l,l'} D_{ll'}^>(t,t') \times \\
 &&{\rm Tr}\left[ M^l G^{>}(t,t') M^{l'} G^a(t',t_1) 
\frac{\partial \Sigma^<(t_1, t_2)}{\partial t_1}  G^r(t_2, t) \right]. \nonumber
\end{eqnarray}
Note the partial derivative on the first argument of $\Sigma^<$, which is represented by a dot in the diagram.
The partial derivative can  be moved around with repeated integration by part.

A key identity \cite{datta-book},
\begin{equation}
G^r( \Sigma^> - \Sigma^< ) G^a = 
G^a( \Sigma^> - \Sigma^< ) G^r = G^r - G^a = G^>  - G^<, 
\end{equation}
 is needed to show that the 10 diagrams cancel and reduce to only two.   Here the self energies are total 
 lead self energy (for layer 1 only).  This identity is a simple consequence of the Dyson equation
 $(G^r)^{-1} = (g_c^r)^{-1} - \Sigma^r$, where $g_c^r$ is the Green's function of isolated center.
 From the above equation we can show that
 \begin{equation}
 G^a \Sigma^{>} G^r = G^{>}  + C,
 \end{equation}
here we define $C = G^a \Sigma^{>,<} G^r - G^r \Sigma^{>,<} G^a$, and is the same for greater and lesser
components.   $C$ is anti-hermitian, $C^\dagger = - C$.  $C=0$ if matrices are actually 1 by 1, or
if system is time-reversal symmetric \cite{zhang2018energy}, but not so in general.
From this, ignoring the proportionality constant, integration variables, and $M$ factors, we can
write, symbolically, 
\begin{equation}
\label{eq-delta3}
\Delta 3 + \Delta 3') = {\rm Tr} \bigl[ (D^> G^> - D^< G^< ) C\bigr].
\end{equation}
Here the notation $\Delta$ means that the term when $G^a$ and $G^r$ are swapped to form $G^>$ or 
$G^<$ has been subtracted off.   We show that Eq.~(\ref{eq-delta3})  cancels all the other 8 diagrams. 
To this end, we define
\begin{equation}
B = G^> \Sigma^< - G^< \Sigma^>.
\end{equation}
Using the same identities, we have $B G^r = - C$, thus $BG^r - G^a B^\dagger = -2C$, and
$B G^r + G^a B^\dagger = 0$. 

We can factor out common factors in the remaining diagrams.  Using $B$, we can write
\begin{eqnarray}
1\!+\!1') + 2 \!+\!2') &=& D^< {\rm Tr}(G^r B G^r) + D^r {\rm Tr}(G^> B G^r), \nonumber \\
4\!+\!4') + 5\! +\!5') &=& D^< {\rm Tr}(G^a D^a B^\dagger) + D^a {\rm Tr}(G^> G^a B^\dagger).\nonumber
\end{eqnarray}
Further simplification is possible because
\begin{equation}
D^< G^r + D^r G^> = D^> G^> - D^< G^< + D^< G^a + D^a G^>. \nonumber 
\end{equation}
Now, putting all the terms together, and using the identities obtained, we see
$\Delta 3 + \Delta 3')$ cancels all the rest as claimed.

The remaining two terms can be transformed into the desired form.  First, we need to move the
derivative to other places, for example, from graph $3  - \Delta 3)$, we can write
\begin{equation}
- D^>(t,t') {\rm Tr}\bigl[ G^>(t,t') {\partial \over \partial t} G^<(t',t) \bigr].
\end{equation}
The extra minus sign is due to the integration by part.   We can combine a similiar term
from $3' - \Delta 3')$ so that it becomes $\partial \Pi^<(t',t)/\partial t$, using integration by part
and cyclic permutation of trace whenever is needed.  Using the definition of polarization
\begin{equation}
\Pi^{<}_{l'l}(t',t) = - i\hbar\, {\rm Tr}\bigl[  M^{l'} G^<(t',t) M^{l} G^{>}(t,t') \bigr], 
\end{equation}
and then Fourier transform
the final expression to frequency domain, we obtain
\begin{equation}
I_1 =- \frac{1}{4\pi} \int_{-\infty}^{+\infty}\!\!
d\omega\, \hbar \omega\, {\rm Tr} \bigl(D^> \Pi^<_1 - D^< \Pi^>_1 \bigr).
\end{equation}
The heat current density is given by $H = I_1/A$ where $A$ is the area of one layer surface.


\bibliography{MWref}

\end{document}